# Improving student understanding of quantum mechanics underlying the Stern-Gerlach experiment using a research-validated multiple-choice question sequence

Paul Justice, Emily Marshman, and Chandralekha Singh

*Department of Physics and Astronomy, University of Pittsburgh, Pittsburgh, PA, 15260*

**Abstract.** Engaging students with well-designed multiple-choice questions during class and asking them to discuss their answers with their peers after each student has contemplated the response individually can be an effective evidence-based active-engagement pedagogy in physics courses. Moreover, validated sequences of multiple-choice questions are more likely to help students build a good knowledge structure of physics than individual multiple-choice questions on various topics. Here we discuss a framework to develop robust sequences of multiple-choice questions and then use the framework for the development, validation and implementation of a sequence of multiple-choice questions focusing on helping students learn quantum mechanics via the Stern-Gerlach experiment that takes advantage of the guided inquiry-based learning sequences in an interactive tutorial on the same topic. The extensive research in developing and validating the multiple-choice question sequence strives to make it effective for students with diverse prior preparation in upper-level undergraduate quantum physics courses. We discuss student performance on assessment task focusing on the Stern-Gerlach experiment after traditional lecture-based instruction vs. after engaging with the research-validated multiple-choice question sequence administered as clicker questions in which students had the opportunity to discuss their responses with their peers.

## I. INTRODUCTION

**Background**: A major goal of many physics courses from introductory to advanced levels is to help students learn physics concepts [1-7] while also helping them develop problem solving and reasoning skills [8-18]. We have been investigating strategies to help students develop a solid grasp of physics concepts and develop their problem solving and reasoning skills [19-31]. In fact, many education researchers have been involved in developing and evaluating evidence-based active-engagement (EBAE) curricula and pedagogies [32-50], but implementation of these EBAE approaches to help college students learn has been slow. Some of the major barriers to implementation of the EBAE pedagogies at the college-level include lack of faculty buy-in and their reluctance and/or resistance, partly due to a lack of institutional reward system for using these evidence-based approaches, the time commitment involved in effectively adapting and implementing them, and instructors' fear that their students may complain (since students may prefer to passively listen to lectures as opposed to actively engage in the learning process) [32]. Moreover, the amount of class time required to implement an EBAE pedagogy, the flexibility with which it can be implemented, the need to train instructors in how to effectively use it, and the architectural constraints of the classrooms may also increase the barrier and make it difficult to implement an EBAE pedagogy [32].

**Use of Multiple-choice questions and peer interaction**: The use of well-designed multiple-choice questions in the classroom in which each student must first take a stand before discussing the responses with peers is an EBAE pedagogy that has a relatively low barrier to implementation. These multiple-choice questions can be used in a variety of ways to engage students actively in the learning process. For example, "think-pair-share" approach in which each student thinks about why a particular answer choice is correct for each question first and then pairs up with a peer sitting next to them to share their thoughts before a general discussion about the correct answer can be an effective pedagogy. The use of multiple-choice questions keeps students focused during the lectures and helps them self-monitor their learning. Moreover, depending upon instructional goals and instructors' preferences, multiple-choice questions can be used in diverse manner, e.g., they can be interspersed within lectures to assess student learning of the material in each segment, posed at the end of a class or used to review materials from previous classes. They can be used in a flipped class in which the class time is entirely focused around multiple-choice questions and discussions around them [32].

Integration of peer interaction with lectures using multiple-choice clicker questions has been made popular by Mazur in the physics community [33]. In Mazur's approach, the instructor poses conceptual, multiple-choice clicker questions to students throughout the class. Students first answer each multiple-choice question individually using clickers (or personal response system), which requires them to take a stance regarding their understanding of the concepts involved [33]. Next, they discuss the questions with their peers and learn by articulating their thought processes and assimilating their ideas and understanding with those of the peers. Then, there is a class discussion involving all students about those concepts in which both students and the instructor participate fully. By having students take a stand anonymously using clickers as opposed to using a show of hand for each answer choice selected for a multiple-choice question, students do not feel embarrassed if their answer choices are not correct. Moreover, clickers offer another advantage over a show of hand or show of cards with different colors (for different answer choices selected for each multiple-choice question) in that the responses are recorded and the instructor can contemplate how to address the common difficulties that students have (which they otherwise may not feel comfortable sharing with the instructor). In particular, the immediate feedback that is obtained by the instructors is valuable because they have an understanding of the extent of student difficulties and the percentage of students who understand the concepts.

**Multiple-choice Question Sequences**: While multiple-choice questions, which can be used with or without clickers, have been developed [33-45] for introductory physics and upper-level physics such as quantum mechanics (QM), there have been very few documented efforts [45] toward a systematic development of multiple-choice question sequences (MQSs) in which the set of questions build on each other effectively and help students extend, organize and repair their knowledge structure. Moreover, in the past two decades, many investigations have focused on improving student learning of QM, e.g., see [51-64]. Our group has been involved in such investigations and we are using the research on student difficulties as a guide to develop research-validated learning tools [65-84]. Our past investigations in upper-level QM courses suggest that while engaging students with multiple-choice questions during class can help them learn, they may not be as effective as a research-validated QuILT on the same concepts unless they are carefully sequenced [68]. For example, we find that when students in upper-level undergraduate QM course only had traditional lecture-based instruction, their performance on a quantum measurement quiz covering those concepts was 26% [68]. The next year, the student performance in the same course on an equivalent quiz on quantum measurement concepts was 68% after lecture and multiple-choice questions on quantum measurement, and their score on another equivalent quiz was 91% after engaging with a QuILT on quantum measurement [68]. A research-validated MQS has the potential to bridge the gap between students' performance after they engage with a QuILT vs. after they engage only with a MQS on the same topic (since the multiple-choice questions in a MQS build on each other and can take advantage of the learning objective and guided sequences in the corresponding QuILT).

Here we discuss a framework for the development, validation and implementation of a MQS and then apply it to develop and validate an MQS to help students learn about Stern-Gerlach experiment, a topic which is valuable for exposing students to the foundational issues in QM, by taking advantage of the guided learning sequences in a research-validated Quantum Interactive Learning Tutorial (QuILT) on that topic [66]. The SGE MQS strives to help students learn fundamental issues in QM using a simple two-state system. We also discuss the implementation of this MQS by two different instructors using clickers with peer discussions interspersed throughout the class. A QuILT focusing on a QM concept uses a guided, inquiry-based approach to learning and consists of learning sequences, which are based both upon a cognitive task analysis from an expert perspective and an extensive research on student difficulties in learning those concepts. The QuILT was useful both for developing new multiple-choice questions and revising/fine-tuning existing multiple-choice questions (e.g., in situations in which individual multiple-choice questions are already validated but the sequencing of questions is not validated), or for developing entirely new multiple-choice questions to ensure that different questions in the MQS build on each other.

Before we focus on a MQS that strives to help students learn about the Stern-Gerlach experiment (SGE), we enumerate the learning objectives.

**Learning objectives of the SGE MQS:** The learning objectives of the SGE MQS are commensurate with the corresponding QuILT [66], which guided the development and sequencing of the SGE MQS questions. These learning objectives are focused on improving student understanding of the foundational issues in QM via the Stern-Gerlach experiment and were developed using extensive research on student difficulties with these concepts and cognitive task analysis from an expert perspective [66]. These foundational issues include the difference between the physical space (in which the experiment is performed) and the Hilbert space (in which the state of the quantum system lies), quantum state preparation, quantum measurement and the difference between a superposition and mixture. We find that after traditional lecture-based instruction, many students have difficulty differentiating between the physical space and Hilbert space. For example, they believe that if a neutral silver atom in an eigenstate $|\uparrow\rangle_x$ of the $x$ component of the spin angular momentum $\hat{S}_x$ is sent through a Stern-Gerlach apparatus (SGA) with magnetic field gradient in the $z$ direction, the magnetic field will not impact the quantum state because the magnetic field gradient is orthogonal to the quantum state $|\uparrow\rangle_x$. This type of reasoning is incorrect because the quantum state is a vector in the Hilbert space in which the state of the system lies whereas the magnetic field is a vector in the physical space in which the experiment is being performed. It does not make sense to talk about orthogonality of two vectors in different vector spaces. Thus, the first learning objective of the MQS is to help students develop a solid grasp of the difference between the physical space and Hilbert space using the SGE with a two state system as an example. The second learning objective focuses on helping students develop a functional understanding of quantum state preparation because one fundamental issue that the SGE can beautifully illustrate at least conceptually using a two state system is the issue of how to prepare a quantum state. For example, if a quantum system consisting of a large number of neutral silver atoms is initially in an eigenstate $|\uparrow\rangle_z$ of the z component of the spin angular momentum $\hat{S}_z$, is it possible to use appropriate SGAs and detectors to prepare a state which is orthogonal to it, i.e., $|\downarrow\rangle_z$? The third learning objective focuses on helping students learn about quantum measurement in the context of a simple two-state system (e.g., using a beam of neutral silver atoms which can be treated as a spin-1/2 system) in a given spin state and how the state of the silver atoms is impacted by passing through a SGA and how the placement of the detectors (that can detect the silver atoms) in appropriate positions will collapse the state to different eigenstates of the observable measured with different probabilities, depending upon the set up. The fourth learning objective is to help students understand that there is a difference in the situations in which a beam of silver atoms propagates through a series of SGAs but no measurement via a detector is performed vs. the case in which a detector is present for measuring the silver atoms deflected upward or downward (because the measurement will collapse the state of the system into an eigenstate of the measured observable with different probabilities). Since the orbital and spin degrees of freedom of the silver atoms are entangled, a detector after a SGA in the up channel or down channel clicking would signify the spin state of the silver atom collapsing to a particular state. The fifth learning objective is to help students be able to analyze the probabilistic outcome of a measurement when a given initial two-state system is sent through a SGA by transforming the initial state given in a basis to another basis that is more suited for the analysis of measurement outcomes based upon the magnetic field gradients (e.g., if the Stern-Gerlach apparatus has a gradient in the $x$ direction, the most convenient basis in which the incoming state should be transformed to analyze the measurement outcomes consists of eigenstates of the x component of the spin angular momentum). Finally, the sixth learning objective of the MQS is to help students be able to develop a functional understanding of the difference between a superposition vs. mixture and how certain experimental configurations involving SGAs are able to differentiate between them.

## II. METHODOLOGY FOR DEVELOPING, VALIDATING AND IMPLEMENTING A MQS

Before discussing the SGE MQS, we first summarize general issues involved in the development, validation and implementation of a robust MQS using the inquiry-based learning sequences in the corresponding QuILT as a guide. In particular, below, we summarize some of these "lessons learned" from the guided sequences in the QuILT that can be used as a guide to develop and/or revise a MQS (first three points below) and implement it effectively (last three points below):

**1. Balance difficulty:** A QuILT is structured such that students are provided enough guidance to develop a coherent conceptual understanding without becoming frustrated or discouraged. Following this principle and earlier suggestions to make effective use of class time, e.g., by Mazur et al. [33], we decided that a majority of the questions in a MQS should have correct response rates such that both extremes are avoided (i.e., we avoided cases in which very few students answer the question correctly or incorrectly). If some students already have a reasonable understanding of the topic, it is likely to make the peer discussions effective and encourage students to engage in productive discussions and learn the concepts with peer support. A question in which very few students can reason about it correctly may result in reinforcing students' inaccurate conceptual models and it also becomes more likely that students would guess as opposed to apply the physics concepts systematically. On the other hand, high scores indicate that there is little constructive struggle. With restricted class time, such questions should be limited except for warmups (to help students review basic concepts and prime them to answer more complex questions later) [45].

**2. Change only the context or the concept between questions:** We took inspiration from the corresponding QuILT [66] on the same topic to ensure that different questions in the MQS build on each other and ascertain how changes in context and concepts should be included in the MQS. We found that switching both the concept and the context in adjacent questions may result in cognitive overload for students. Changing only the context or concept between consecutive questions may help students identify the differences and similarities between subsequent questions and construct correct models more effectively.

**3. Include a mix of abstract and concrete questions:** Although the type of questions in a MQS is usually dictated by the topic and the goals and learning objectives, we examined guided learning sequences in the QuILT [66] to determine how to pose abstract and concrete questions in a MQS. Abstract questions may provide students opportunities to generalize concepts across different contexts. On the other hand, concrete questions allow students to apply their learning to a concrete context. Students may benefit from a balance of both question types.

**4. Allow student collaboration:** Collaborative group work is found to be beneficial for helping many students learn [33-50]. A student who is having difficulty and a student who is articulating her thoughts both refine their understanding so that co-construction of knowledge can occur when neither student was able to answer the questions before peer collaboration, but were able to answer them correctly after working together [33]. Thus, instructors should allow for peer discussion while implementing a MQS.

**5. Incorporate "checkpoints" at appropriate times during a MQS:** A QuILT often includes checkpoints which provide opportunity to reconcile the differences between student ideas and the correct conceptual model. The MQS for a given topic can include "checkpoints" at similar points as the QuILT, at which the instructor can have a general class discussion and can give feedback to the entire class based upon students' multiple-choice question responses to help them learn.

**6. Include a manageable number of multiple-choice questions per sequence:** The researchers deliberated and concluded (based upon data from multiple-choice questions in previous years and the learning sequences in the QuILT) that a MQS should include a manageable number of questions that should build on student prior knowledge. Having many questions in a sequence may offer students more opportunities to practice concepts, but having too many questions can result in a sequence of multiple-choice questions that cannot be reasonably implemented effectively, given the time constraints of the class.

## III. METHODOLOGY FOR DEVELOPING AND VALIDATING SGE MQS

In order to develop and validate effective sequences of multiple-choice questions for QM focusing on the SGE, three researchers met to holistically examine the instructional materials from the past few years on these topics in an upper-level undergraduate QM course at a large research university in the US, which included existing multiple-choice questions and the QuILT on this topic. This course typically has 15-25 students each year, who are mainly physics juniors/seniors. The validation of MQS was an iterative process. Moreover, the questions in the SGE MQS were developed or adapted from prior validated multiple-choice questions and sequenced to balance difficulties, avoid change of both concept and context between adjacent questions as appropriate in order to avoid experiencing cognitive overload, and include a mix of abstract and concrete questions to help students develop a good grasp of the concepts. In particular, in order to design an effective SGE MQS, we examined the SGE QuILT [66] and contemplated how to take advantage of its learning objectives, guided learning sequences, and student performance on the pre-/posttests administered before and after they engaged with it.

While developing the SGE MQS, we drew upon the learning objectives delineated earlier and the requisite knowledge and skills required to achieve those objectives. We also focused on the order in which different QM concepts that are involved in the learning objectives are invoked and applied in a given situation within the SGE QuILT to inform the design of the SGE MQS. Furthermore, we carefully examined the types of scaffolding provided in the SGE QuILT [66] to reflect on how different questions within the SGE MQS should effectively build on each other and whether more scaffolding between some existing multiple-choice questions is required. We also analyzed student performance on SGE multiple-choice questions in previous years to determine whether students were able to transfer their learning from one multiple-choice question to another and whether some multiple-choice questions would require more scaffolding between them in order to be effective. In particular, when examining the SGE MQS that had been administered in the previous years and comparing them with the guided learning sequences in the SGE QuILT, we realized that sometimes both the concept and the context changed from the preceding to the following question in an old sequence. We hypothesized that this may cause cognitive overload for students and at least partly be responsible for making the previous set of multiple-choice questions less effective (for which we had evidence from the data from previous years). We took inspiration from the SGE QuILT to develop the set of questions for the SGE MQS that have appropriate ordering and balance to scaffold student learning and help them compare and contrast different concepts and contexts effectively. The issue of abstract vs. concrete questions was also deliberated. Abstract questions posed tend to focus on generalized cases whereas concrete questions, in general, involve a specific context. It was decided that only the last question in the MQS will have an abstract choice since the SGE is best learned using concrete examples using diverse setups of SGAs and initial states.

After the initial development of the SGE MQS using the learning objectives, inquiry-based guided sequences in the QuILT and existing individually validated questions, we iterated the MQS with three physics faculty members who provided valuable feedback. The feedback from faculty helped in fine-tuning and refining some new questions that were developed and integrated with the existing ones to construct the sequence of questions in the SGE MQS and to ensure that the questions were unambiguously worded and build on each other based upon the learning objectives. We also interviewed four students individually who answered the MQS questions in a one on one interview situation while thinking aloud so that we could understand their thought processes. The four interviews totaled about 3 hours. These interviews, which also reaffirmed the common difficulties earlier described by Zhu et al. [66], ensured that students found these questions unambiguous and were able to take advantage of the scaffolding provided by different questions that build on each other. These student interviews were helpful for further tweaking the questions.

## IV. MQS FOCUSING ON THE SGE THAT WERE IMPLEMENTED IN CLASS

After the out-of-class development and validation, the final version of the SGE MQS that went through in-class implementation via clickers along with peer discussion has 7 questions (MQ1-MQ7). As discussed in the next section, two different instructors at the same institution implemented the SGE MQS in two consecutive years in the upper-level undergraduate QM course such that the pretest was given after traditional lecture-based instruction, and then students engaged with the entire SGE MQS with 7 questions in class before they were administered the posttest. There was no overall class discussion after MQ1 but there was an overall class discussion after each of the other questions in the SGE MQS.

Similar to the QuILT [66], in the MQS administered as clicker questions with peer interaction and the corresponding pretest and posttest, the description of the Stern-Gerlach apparatus shown in Figure 1 was provided to students because it is important to clarify the notation used for the SGAs. Students also knew that the orbital angular momentum of a beam of neutral silver atoms is zero, so they had to focus on the fact that a beam of neutral silver atoms passing through a SGA can be considered a spin-1/2 system. They had learned that a SGA can entangle the orbital and spin degrees of freedom depending upon the initial state of the system and the SGA setup. Students also were asked to assume that the detectors were placed in appropriate orientations after a SGA and when a detector clicks, the silver atom is absorbed by that detector.

The figure below shows the pictorial representations used for a Stern-Gerlach apparatus (SGA). If an atom with state $|\uparrow\rangle_z$ (or $|\downarrow\rangle_z$) passes through a Stern-Gerlach apparatus with the field gradient in the negative z-direction (SGZ-), it will be deflected in the +z (or -z) direction. If an atom with state $|\uparrow\rangle_z$ (or $|\downarrow\rangle_z$) passes through a Stern-Gerlach apparatus with the field gradient in the positive z-direction (SGZ+), it will be deflected in the -z (or +z) direction. Similarly, if an atom with state $|\uparrow\rangle_x$ passes through SGX- (or SGX+), it will be deflected in the +x (or -x) direction. The figures below show examples of deflections through the SGX and SGZ in the plane of the paper. However, note that the deflection through a SGX will be in a plane perpendicular to the deflection through an SGZ. This actual three-dimensional nature should be kept in mind in answering the questions.

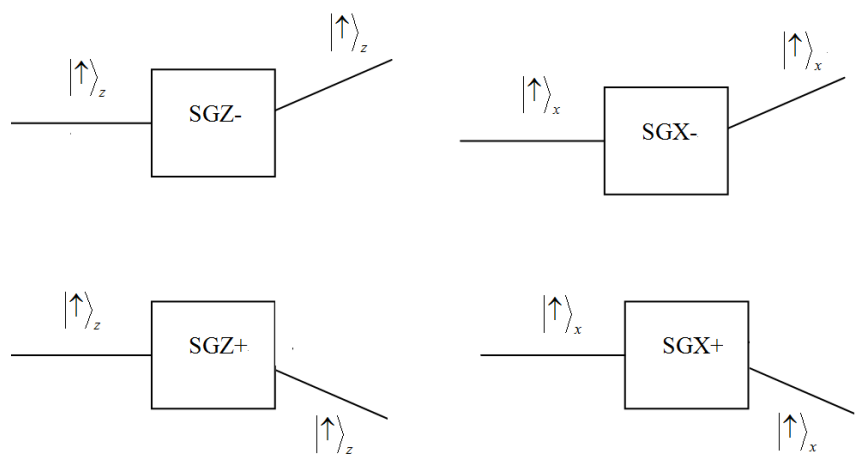

**Figure 1.** This information is provided to students in all contexts (e.g., before the MQS and with the pretest and posttest).

The first two questions of the MQS, questions MQ1 and MQ2, focus on student difficulty in differentiating between Hilbert space and physical space as well as on the choice of an appropriate basis to analyze the probability of measuring different outcomes given a particular initial state of the system and the SGE setup as follows. Correct answers are in bold for all multiple-choice questions.

**(MQ1)** A beam of neutral silver atoms in a spin state $|\chi\rangle = \frac{1}{\sqrt{2}}(|\uparrow\rangle_z + |\downarrow\rangle_z)$ propagates into the screen (x-direction) as shown in Figure 2. The beam is sent through a SGE with a horizontal magnetic field gradient in the –z-direction. What is the pattern you predict to observe on a distant screen in the y-z plane when the atoms hit the screen?

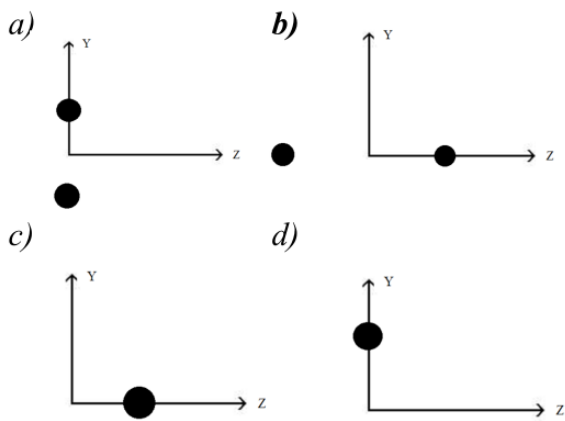

**Figure 2.** Figure for MQ1.

**(MQ2)** A beam of neutral silver atoms in a spin state $|\chi\rangle = \frac{1}{\sqrt{2}}(|\uparrow\rangle_z + |\downarrow\rangle_z)$ propagates into the screen (x-direction) as shown in Figure 3. The beam is sent through a SGE with a horizontal magnetic field gradient in the –y-direction. What is the pattern you predict to observe on a distant screen in the y-z plane when the atoms hit the screen?

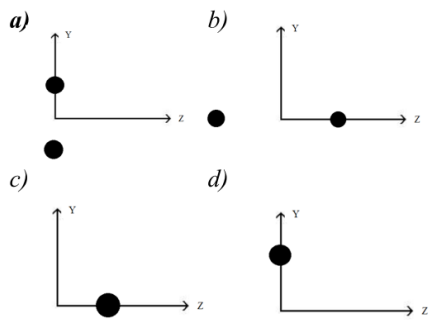

**Figure 3.** Figure for MQ2.

In the in-class administration discussed in the next section, after students answered both questions using clickers after discussing their responses with a peer sitting next to them, a whole class discussion led by the instructor focused on common student difficulties, e.g., in differentiating between physical space and Hilbert space, the importance of choosing an appropriate basis and transforming the initial state in that basis in order to analyze the measurement outcomes before posing MQ3 as a clicker question. In question MQ3 (see Appendix II), students were told that $|a|^2+|b|^2=1$. In MQ3, silver atoms in a generic spin state given in the basis of eigenstates of $\hat{S}_z$ are sent through a single Stern-Gerlach apparatus with an $x$ gradient. This question focuses on helping students learn about the usefulness of transforming from the given basis to a more suitable basis in order to analyze the measurement outcomes including the probability of the detector clicking. It also helps students learn about preparing a specific spin state, here $|\downarrow\rangle_x$. Building on this question, MQ4-MQ6 (see Appendix II) ask students to contemplate issues related to preparation of quantum states in different arrangements of SGAs and initial state. In MQ5, students were told to assume that the strength of the second SGA was such that any spatial separation of the state after the first SGA was negated. Finally, MQ7 focuses on helping students think about how to use the SGE to differentiate between a state which is a superposition of eigenstates of an operator corresponding to an observable and a mixture. We note that oblique lines are shown as a guide only in some of the figures (e.g., Figures 5, 6 and 7) as a scaffolding support but not shown in others, e.g., in Figures 4 and 9 so that students learn to think about and draw them themselves. We also note that there are many experimental considerations in constructing a real SGA, e.g., the challenge of maintaining coherence between two spatially separated beams such as in MQ5. These considerations were not discussed by the instructor (since it has the possibility to increase student cognitive load) but would certainly add richness to the topic.

## V. METHODOLOGY FOR IN-CLASS IMPLEMENTATION

Not only did we focus on developing and validating the SGE MQS, we also contemplated effective strategies for their in-class implementation. For example, we used the SGE MQS with clickers and peer discussion, a principle emphasized by Mazur [33]. In fact, empirical data from QM multiple-choice questions given in previous years suggest that, on average, student performance on multiple-choice questions improved significantly after discussing responses with a peer [35]. We also considered the need for productive struggle for students when working on the SGE MQS. For example, we considered the points at which the instructor should provide feedback to students in order to maximize productive engagement and minimize discouragement. Again we drew both upon the SGE QuILT and empirical data from student responses to multiple-choice questions in previous years to identify where feedback might be most effective in the SGE MQS. We paid attention to the QuILT "checkpoints" for students to resolve possible conflicts between their understanding and the correct conceptual model of the foundational issues elucidated via the SGE [66]. These "checkpoints" guided us in identifying points for instructor feedback and general class discussion for the SGE MQS that may be beneficial for students. We also analyzed the points at which students struggled when answering clicker questions in previous years to determine possible points during the implementation of the SGE MQS when a particular general class discussion was likely to be effective. It was determined that one productive approach would be for the instructor to have students answer the first two SGE MQS questions before having a general class discussion followed by a class discussion after each of the following SGE MQS questions.

We note that the final version of the 7 SGE MQS questions can be integrated with lectures in which these relevant concepts pertaining to SGE are covered in a variety of ways based upon the instructor's preferences. For example, the MQS can be given over multiple sessions or together depending, e.g., upon whether these are integrated with lectures, used at the end of each class, or used to review concepts after students have learned about the Stern-Gerlach experiment via lectures. However, in this study, the SGE MQS was implemented with clickers and peer discussion [33] in an upper-level undergraduate QM class at a large research university after traditional lecture-based instruction in relevant concepts on the Stern-Gerlach experiment for two consecutive years by two different instructors. Both instructors tried their best to implement the SGE MQS in a very similar manner using clickers and peer discussion and with similar general whole class discussions that was deemed effective as discussed in the preceding paragraph. Prior to the implementation of the SGE MQS in class, students were administered a pretest after traditional lecture-based instruction, which was developed and validated by Zhu et al. [66] to measure comprehension of the relevant concepts. The students then engaged in class with the SGE MQS and discussed their answers with their peers. This implementation was completed in one class period. The posttest was administered during the following week to measure the impact of the SGE MQS on student learning of relevant concepts.

The posttest that students were administered following the implementation of the SGE MQS was analogous to the pretest [66]. These pre-/posttests are the same as those administered by Zhu et al. to measure student learning after traditional lecture-based instruction and after engaging with the SGE QuILT [66]. In order to compare the performance of the SGE MQS and QuILT groups on pre-/posttests so that the relative improvements can be determined, the same rubric was used for the pre-/posttests given to the SGE MQS students as the corresponding QuILT students in Ref. [66] (who were also advanced undergraduate students in QM course at the same university). All questions were scored out of a possible 2 points, with partial credit assigned to answers that were correct, but for which either incorrect justification or no justification was provided if reasoning was requested. The inter-rater reliability was better than 95%.

We also note that two versions (test versions A and B) of the tests were designed to be administered as a pretest or posttest. In particular, all questions on the two versions of the test are not identical because we wanted to investigate how students answer questions when the pretest and posttest questions are the same or different. Moreover, the version of the test that was used as a pretest before the SGE MQS in one year was used as a posttest in the other year. The first two questions in the pretest and posttest are analogous to MQ1 and MQ2 in the SGE MQS. Question 1 on both versions is identical and pertains to a state, which is an equal superposition (no relative phase factor) of the spin-up and spin-down states in the z-basis, passing through a SGE with a magnetic field gradient in the -z direction (identical to MQ1). Question 2 has silver atoms with spin-up in the z-basis passing through a SGE with a magnetic field gradient in either the -y direction (as in MQ2) or –x direction, depending on the version of the test. These first two questions address the first learning objective. Question 3 on both versions of the test corresponds to MQ3. On both versions A and B of the test (see Appendix III and Figure 13), question 3 asks about the measurement outcomes for a superposition of spin-up and spin-down states in the z-basis passing through a SGX-. On version A, both coefficients of the state are given numerically, whereas they are given in terms of complex numbers "a" and "b" in version B with $|a|^2+|b|^2=1$. This question addresses both the third and fifth learning objectives.

Question 4 on both versions most closely matches MQ4, but also connects with MQ5 and MQ6. The question 4 in version B is shown in Figure 9 in Appendix III. On version A, the first SGE seen in MQ4 is a SGY-, rather than SGX-. On version B, there is an additional SGZ- in front of the SGX-. This question emphasizes the second learning objective, while also addressing the third, fourth, and fifth learning objectives.

Question 5 on test version A and the analogous test question on version B align with the learning objectives underlying MQ7, i.e., they assess whether students are able to differentiate between a superposition state and an analogous mixture using SGAs. In both versions (see Appendix III), students are given two beams of silver atoms: a superposition $|\chi\rangle = \sqrt{\frac{1}{2}}|\uparrow\rangle_z + \sqrt{\frac{1}{2}}|\downarrow\rangle_z$, and the analogous mixture of 50% $|\uparrow\rangle_z$ and 50% $|\downarrow\rangle_z$. Version A asks students to design a setup of SGAs to differentiate between these two beams (superposition and mixture), while version B asks students to identify which combination of three statements is true regarding these beams passing through different SGAs (see Appendix III). This question emphasizes the sixth learning goal, while also addressing the third and fifth.

Finally, on the same topic as question 4, question 6 (see Appendix III) on both versions of the test assesses student understanding of preparing a state (here students are given an initial state and asked if they can prepare a given orthogonal state using SGAs and how they may be able to do that), and connects most closely with MQ6 asking students to design a setup of SGAs to prepare a spin-up state in the *z*-basis given an initial spin-down state in the *z*-basis. This question emphasizes the second learning objective, while also addressing the third, fourth, and fifth.

## VI. IN-CLASS IMPLEMENTATION RESULTS

In one year of the SGE MQS implementation in upper-level undergraduate QM, students were administered version A as a pretest and version B as a posttest, while in the other year, they were administered version B as a pretest and version A as a posttest. On the other hand, when the SGE QuILT was implemented [66], some of the students in the same class were given version A as the pretest whereas others were given version B (and the versions were switched for each student for the posttest). Tables 1 and 2 compare pre-/posttest performances of students in upper-level QM course from the same university in different years after traditional lecture-based instruction (pretest) and on the posttest after students had engaged with the SGE MQS (Table 1) or SGE QuILT (Table 2). The normalized gain (or gain) is calculated as $g = (post\% - pre\%)/(100\% - pre\%)$

[85]. Effect size was calculated as Cohen's $d = (\mu_{post} - \mu_{pre})/\sigma_{pooled}$ where $\mu_i$ is the mean of group $i$ and where the pooled standard deviation (in terms of the standard deviations of the pre- and posttests) is $\sigma_{pooled} = \sqrt{\sigma_{pre}^2 + \sigma_{post}^2}$ [85]. Normalized gain and effect size are only shown in Table 1 (not available for Table 2 data in Ref [66]).

**Table 1.** Comparison of the mean pre-/posttest scores on each question, normalized gains and effect sizes for students in upper-level undergraduate QM (averaged over two years in which the corresponding questions in versions A and B are averaged) who engaged with the SGE MQS (N=48).

| Question | Pretest Mean | Posttest Mean | Normalized Gain (g) | Effect Size (d) |
|---|---|---|---|---|
| 1 | 61% | 96% | 0.88 | 0.50 |
| 2 | 40% | 92% | 0.87 | 0.71 |
| 3 | 38% | 49% | 0.18 | 0.13 |
| 4 | 51% | 80% | 0.59 | 0.38 |
| 5 | 24% | 50% | 0.34 | 0.28 |
| 6 | 43% | 82% | 0.69 | 0.46 |

**Table 2.** Comparison of mean pre-/posttest scores on each question from Ref. [66] (effect sizes not available) for students in upper-level undergraduate QM who engaged with the SGE QuILT. Questions from versions A and B were mixed in both pre- and posttest in that some students got version A as the pretest and others as the posttest (and vice versa). Mean scores are not for matched students and numbers of students varies from 5 to 35 (more details can be found in Ref.[66]).

| Question | Pretest Mean | Posttest Mean |
|---|---|---|
| 1 | 80% | 81% |
| 2 | 39% | 77% |
| 3 | 30% | 80% |
| 4 | 40% | 90% |
| 5 | 42% | 90% |
| 6 | 38% | 100% |

Tables 1 and 2 show that students in general did not perform well on the pretest after traditional lecture-based instruction. Note that we have combined the data for questions which have been treated as equivalent between versions A and B, whereas Zhu et al. refer to questions from different versions separately in Ref.[66].

One common difficulty on the pretest, illustrated in Figure 10, is the belief that the state $|\uparrow\rangle_z$ of the beam propagating through the Stern-Gerlach apparatus will be deflected in the $z$ direction in response to question 2 on version B of the test.

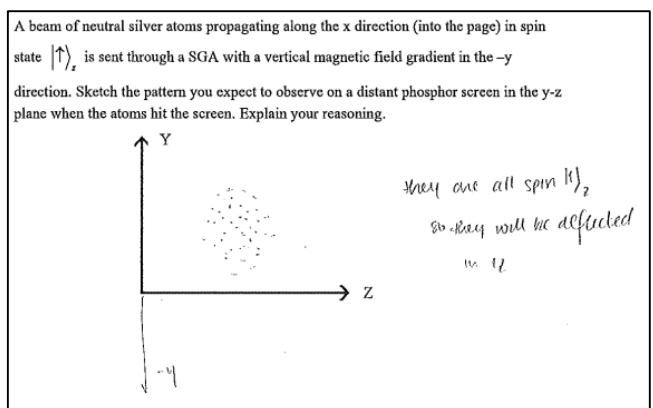

**Figure 10.** An example of a common student belief that the state of the beam propagating through the Stern-Gerlach apparatus will be deflected in the $z$ direction because the state is $|\uparrow\rangle_z$ in response to question 2 on version B of the test.

However, Table 1 shows that there is significant improvement on the first two questions from the pretest (after traditional lecture-based instruction in relevant concepts) to posttest (after the SGE MQS) with students using the SGE MQS scoring greater than 90% on both questions. This improvement in students' understanding is partly due to students being able to differentiate between the Hilbert space and physical space better and also being able to choose a suitable basis for the initial state of the system sent through a SGA and analyze the outcomes of an experiment based upon the magnetic field gradient of the SGA after engaging with the SGE MQS.

Switching between bases and preparation of a state orthogonal to an initial state are investigated via test questions 4 and 6 (see Figures 3 and 4 for version B) and MQ4. Table 1 also shows that on both questions 4 and 6, there was a reasonable gain from pretest to posttest after the SGE MQS implementation. While not as large as the gains observed for these questions from pretest to posttest for students using the SGE QuILT (see Table 2), Table 1 shows that students scored approximately 80% in response to these questions after engaging with the SGE MQS (these questions were about a situation in which the final prepared spin state of the silver atoms was a "flipped" state orthogonal to the spin state of the incoming atoms, and a situation in which students had to design an experiment with a series of SGAs with the goal of "flipping" the incoming quantum state).

In contrast, questions 3 and 5 both demonstrate some room for improvement on posttest even after the SGE MQS implementation (see Table 1) compared to the corresponding improvement after the SGE QuILT (see Table 2). An example of a student response to Question 3 on the pretest on version A is shown in Figure 11, in which the student had difficulty recognizing that the initial state is an eigenstate $|\uparrow\rangle_x$ of $\hat{S}_x$ so the entire beam will be deflected upward in this situation. The student stated that the probability of the detector clicking when an atom exits the SGX- is *"50% probability because the SGE is in the x-direction, so it will not deflect particles with z but these particles also have x-components, we just don't know them."* It appears that this student is aware of the fact that "these particles also have x-components" but the student does not know how to find them. Many other students had similar difficulties on the pretest. Another difficulty students had with the orthogonality of states was mistakenly assuming that a spin-up state $|\uparrow\rangle_z$ is orthogonal to the spin-up state $|\uparrow\rangle_x$.

**Figure 11.** An example of a student response to Question 3 on test version A in which the student had difficulty recognizing that the initial state is an eigenstate of $\hat{S}_x$ so the entire beam will be deflected upward. The student states that the probability of the detector will click when an atom exits the SGX- is *"50% probability because the SGE is in the x-direction, so it will not deflect particles with z but these particles also have x-components, we just don't know them."*

Based on our results, the topic of different bases in question 3 on the pretest and posttest is an area in which the SGE MQS should be refined to improve student understanding. Moreover, Table 3 (see Appendix I) shows that students who obtained version A of the test as a pretest and version B of the test as a posttest showed moderate gains from the implementation of the SGE MQS, but the same did not hold true for those who obtained version B as a pretest and version A as a posttest (see Table 4 in Appendix I). Apart from differences between students and instructor implementation in two consecutive years (even though each instructor implemented them using similar approaches to the best of their abilities, there still may be individual differences), one issue that may contribute to the difference between question 3 performances of these two classes in Tables 3 and 4 is that students in Table 4 received this question as a multiple-choice question on version B on their pretest and as an open-ended version A on their posttest shown in Appendix III. In Table 3, students had the reverse situation with regard to the versions. Tables 3 and 4 suggest that the class assessed using version B performed better on this question on the posttest. We also note that the differences in pretest averages before engaging with MQS can likely be due to differences in instructors' lecturing styles and differences between students in two consecutive years, something we do not have any control over. Therefore, we do not want to dwell on the pretest differences on any question in Tables 3 and 4. However, regardless of the pretest performance in Tables 3 and 4, the posttest performance on both versions of question 3 (see Tables 3 and 4 in Appendix I) after the SGE MQS shows room for improvement with regard to helping students learn to transform from one basis to another to analyze measurement outcomes after passing through a SGA with a particular magnetic field gradient. We are contemplating adding another question to provide additional coaching and scaffolding to students in order to solidify their conceptual understanding of how to transform from one basis to another depending upon the magnetic field grading in the Stern Gerlach apparatus and to help students reason about the outcomes of measurement after the atoms pass through a SGA.

Question 5 assesses student proficiency in differentiating between a mixture and superposition of states and showed weak improvement after implementation of the MQS (see Tables 1, 3 and 4). Even after engaging with the MQS, which strived to help students learn to differentiate between a superposition state, $|\chi\rangle = \sqrt{\frac{7}{10}}|\uparrow\rangle_z + \sqrt{\frac{3}{10}}|\downarrow\rangle_z$ and an analogous mixture made up of 70% $|\uparrow\rangle_z$ particles and 30% $|\downarrow\rangle_z$ particles, students struggled with this concept. Previously, students using the SGE QuILT had shown much stronger gains (Table 2). We are currently considering adding another multiple-choice question in the SGE MQS to have students further reflect upon the difference between a superposition of states and a mixture.

## VII. SUMMARY

Well-designed multiple-choice questions with peer discussions are excellent tools for engaging students in the learning process and relatively easy to implement in the classroom, with or without the use of clickers, compared to many other evidence-based active-engagement pedagogies. We describe a framework for developing and validating multiple-choice question sequences and the development, validation and in-class implementation of a MQS focusing on the fundamental concepts in quantum mechanics using the Stern-Gerlach experiment that was inspired by the learning objectives and guided learning sequences in the corresponding QuILT [66]. The SGE MQS was developed using research on student difficulties in learning these fundamental concepts of quantum mechanics as a guide. Different questions in the MQS build on each other and strive to help students organize, extend and repair their knowledge structure. One useful aspect of the Stern Gerlach experiment is that it can help students learn about foundational issues in quantum mechanics using a very simple two state model. In particular, the MQS focuses on helping students learn about important issues in quantum mechanics such as the difference between the Hilbert space and physical space, how to prepare a quantum state, how to analyze the outcomes of a particular set up involving various Stern-Gerlach devices and an initial spin state of neutral silver atoms, and the difference between a superposition state vs. a mixture (and how the SGAs with appropriate orientations of magnetic field gradients can be used to differentiate between these). This MQS is composed of seven questions most of which are posed in concrete contexts with different initial spin states of a beam of neutral silver atoms sent through various SGAs. Only the last question, which focuses on helping students differentiate between a superposition and mixture, concerns an abstract case in which students are asked for the outcome in a situation for which they must consider more than one possible setup to answer correctly. The entire MQS can be spread across separate lecture periods, or can be implemented together, e.g., to review the concepts.

Development of a research-validated learning tool such as the SGE MQS described here is an iterative process. After the in-class implementation of the SGE MQS using clickers and peer interaction by two different instructors, we found that the MQS was effective in helping students learn many of the important concepts. However, in-class evaluation also shows that further scaffolding is needed to guide students in differentiating between a quantum state which is a superposition of eigenstates of an operator corresponding to an observable from a mixture. Appropriate modifications are being made to the SGE MQS so that this issue can be addressed in the future iterations and implementations. Moreover, while both instructors implemented the MQS by interspersing them with lectures using clickers and peer interaction, future research can evaluate the effectiveness of these validated MQS in other modes of classroom implementations.

## VIII. ACKNOWLEDGEMENTS

We thank the National Science Foundation for award PHY-1806691 as well as thank the faculty and students for help.

**APPENDIX I: INDIVIDUAL CLASS DATA**

**Table 3.** Comparison of mean pre-/posttest scores on each question, normalized gains and effect sizes for upper-level undergraduate students in QM who engaged with the SGE MQS when version A was used for pretest and version B was used for posttest (total number of students N=17).

| Question | Pretest Mean | Posttest Mean | Normalized Gain (g) | Effect Size (d) |
|---|---|---|---|---|
| 1 | 38% | 100% | 1.00 | 1.05 |
| 2 | 14% | 97% | 0.97 | 1.79 |
| 3 | 41% | 65% | 0.40 | 0.24 |
| 4 | 38% | 79% | 0.67 | 0.51 |
| 5 | 12% | 59% | 0.53 | 0.64 |
| 6 | 41% | 74% | 0.55 | 0.34 |

Table 4. Comparison of mean pre-/posttest scores on each question, normalized gains and effect sizes for upper-level undergraduate students in QM who engaged with the SGE MQS when version B was used for pretest and version A was used for posttest (total number of students N=31).

| Question | Pretest Mean | Posttest Mean | Normalized Gain (g) | Effect Size (d) |
|---|---|---|---|---|
| 1 | 74% | 93% | 0.72 | 0.28 |
| 2 | 53% | 89% | 0.77 | 0.45 |
| 3 | 35% | 41% | 0.09 | 0.06 |
| 4 | 58% | 80% | 0.53 | 0.30 |
| 5 | 31% | 43% | 0.18 | 0.14 |
| 6 | 44% | 88% | 0.78 | 0.55 |

## APPENDIX II: ADDITIONAL MQS QUESTIONS

**(MQ3)** *A beam of neutral silver atoms in a spin state $|\chi\rangle = a|\uparrow\rangle_z + b|\downarrow\rangle_z$ is sent through a SGX-. An "up" detector blocks some silver atoms, as shown in Figure 4. What fraction of the initial atoms will be blocked by the detector?*

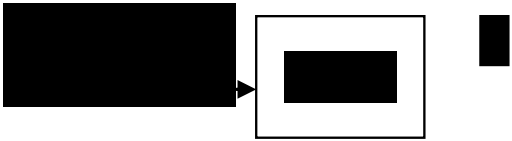

Figure 4. Figure for MQ3 using the representations as described in Figure 1.

a) $|a|^2$   b) $|b|^2$   c) $\frac{1}{2}|a+b|^2$
d) $\frac{1}{2}|a-b|^2$   e) None of the above

**(MQ4)** *A beam of neutral silver atoms is in the initial spin state $|\uparrow\rangle_z$. It propagates through two SGAs as shown in Figure 5. What is the probability that detector B will click for the atoms that enter the first SGA?*

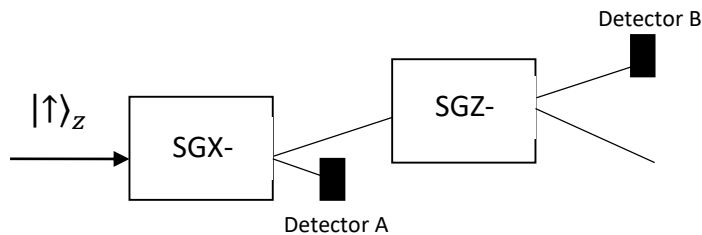

Figure 5. Figure for MQ4 using the representations as described in Figure 1.

a) $1/2$  b) $1/4$  c) $1/8$
d) 1    e) None of the above

**(MQ5)** The initial state of a beam of neutral silver atoms is $|\uparrow\rangle_z$. It propagates through three SGEs as shown in Figure 6. What is the probability that the detector will click for the atoms that enter the first SGE?

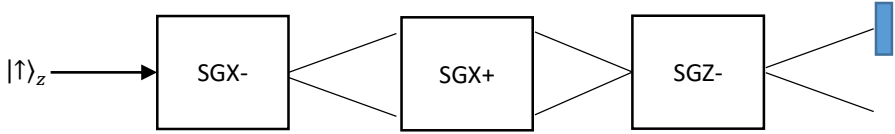

**Figure 6.** Figure for MQ5 using the representations as described in Figure 1.

a) $1/2$  b) $1/4$  c) $1/8$
d) 1    e) None of the above

**(MQ6)** The initial state of a beam of neutral silver atoms is $|\uparrow\rangle_z$. Suppose you want to prepare a beam of neutral silver atoms in spin state $|\downarrow\rangle_z$. Which of the options in Figure 7 shows an appropriate SGE to collect neutral silver atoms (not intercepted by a detector) in spin state $|\downarrow\rangle_z$?

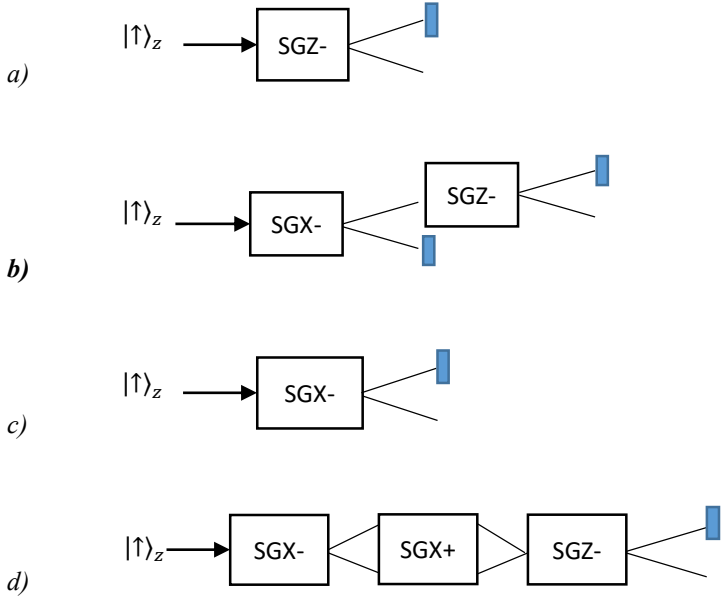

**Figure 7.** Figure for MQ6 using the representations as described in Figure 1.

**(MQ7)** Suppose neutral silver atoms are in an unknown spin state. The spin state is either a mixture with 70% of the atoms in the $|\uparrow\rangle_z$ state and 30% in the $|\downarrow\rangle_z$ state or it is a superposition state $|\chi\rangle = \sqrt{\frac{7}{10}}|\uparrow\rangle_z + \sqrt{\frac{3}{10}}|\downarrow\rangle_z$. Choose all of the following states that are correct about the beam propagating through an SGZ or SGX apparatus:
I. When the beam propagates through the SGZ, 70% of the atoms will register in one detector and 30% of the atoms will register in the other, regardless of the two possibilities for the state.
II. When the beam propagates through the SGX, 50% of the atoms will register in one detector and 50% of the atoms will register in the other, regardless of the two possibilities for the state.
III. We can use a SGZ to distinguish between the possible spin states of the incoming silver atoms.
 a) I only       b) II only
 c) III only     d) II and III only
 e) None of the above

## APPENDIX III: ADDITIONAL TEST QUESTIONS

3. (version B) Harry sends silver atoms all in the normalized spin state $|\chi(t = 0)\rangle = a|\uparrow\rangle_z + b|\downarrow\rangle_z$ through

a SGX-. He places an "up" detector as shown to block some silver atoms and collects the atoms coming out in the "lower channel" for a second experiment. What fraction of the initial silver atoms will be available for his second experiment? What is the spin state prepared for the second experiment? Show your work.

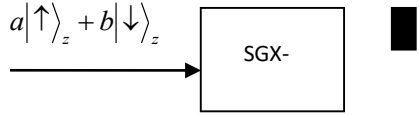

**Figure 8.** Figure for Question 3 on version B that focuses on the learning objective related to transforming the initial state to a basis that makes the analysis of measurement outcomes after passing through the SGA convenient (as well as on how to determine the spin state that is prepared and the fraction of the atoms that are in that final state prepared (i.e., not intercepted by the detector).

4. (version B) Sally sends silver atoms in state $|\uparrow\rangle_z$ through three SGAs as shown below. A detector is placed either in the up or down channel after each SGA as shown. Note that each SGA has its magnetic field gradient in a different direction. Next to each detector, write down the probability that the detector clicks. The probability for the clicking of a detector refers to the probability that a particle entering the **first** SGA reaches that detector. Also, after each SGA, write the spin state Sally has prepared. Explain.

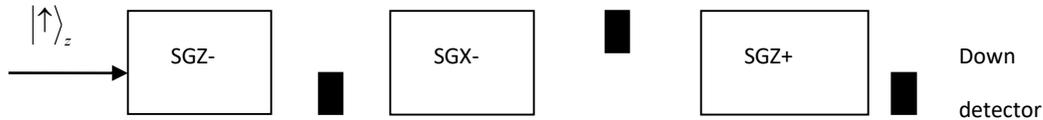

**Figure 9.** Figure for Question 4 on version B that corresponds to the learning objective related to preparing a final quantum state which is orthogonal to the initial state. On both versions of this test question, students were given an arrangement of Stern-Gerlach apparati and were asked to determine the probability that each detector clicks and the spin state prepared.

5 (version A) Suppose beam A consists of silver atoms in the state $|\chi\rangle = \frac{1}{\sqrt{2}}(|\uparrow\rangle_z + |\downarrow\rangle_z)$, and beam B consists of an unpolarized mixture in which half of the silver atoms are in state $|\uparrow\rangle_z$ and half are in state $|\downarrow\rangle_z$. Design an experiment with **SGA**s and **detectors** to differentiate these two beams. Sketch your experiment setup below and explain how it works.

5. (version B) Suppose beam A consists of silver atoms in the state $|\chi\rangle = \frac{1}{\sqrt{2}}(|\uparrow\rangle_z + |\downarrow\rangle_z)$, and beam B consists of an unpolarized mixture in which half of the silver atoms are in state $|\uparrow\rangle_z$ and half are in state $|\downarrow\rangle_z$. Choose all of the following statements that are correct.
    (1) Beam A will <u>not</u> separate after passing through **SGZ-**.
    (2) Beam B will split into two parts after passing through **SGZ-**.
    (3) We can distinguish between beams A and B by passing each of them through a **SGX-**.
A. only 1
B. only 2
C. 1 and 2
D. ***2 and 3***
E. All of the above.
6. (version B) Suppose you have a beam of atoms in the spin state $|\chi(0)\rangle = |\downarrow\rangle_z$ but you need to prepare the spin state $|\uparrow\rangle_z$ for your experiment. Could you use SGAs and detectors to prepare the spin state $|\uparrow\rangle_z$? If yes, sketch your setup below and explain how it works. If no, explain why.